\documentclass[prd,aps,nofootinbib,preprint,preprintnumbers]{revtex4}
\usepackage{graphicx}

\begin{document}

\newcommand{\gsim}{ \mathop{}_{\textstyle \sim}^{\textstyle >} }
\newcommand{\lsim}{ \mathop{}_{\textstyle \sim}^{\textstyle <} }
\newcommand{\vev}[1]{ \left\langle {#1} \right\rangle }

\newcommand{\ds}{\displaystyle}
\newcommand{\bear}{\begin{array}}  \newcommand{\eear}{\end{array}}
\newcommand{\bea}{\begin{eqnarray}}  \newcommand{\eea}{\end{eqnarray}}
\newcommand{\beq}{\begin{equation}}  \newcommand{\eeq}{\end{equation}}
\newcommand{\bef}{\begin{figure}}  \newcommand{\eef}{\end{figure}}
\newcommand{\bec}{\begin{center}}  \newcommand{\eec}{\end{center}}
\newcommand{\non}{\nonumber}  \newcommand{\eqn}[1]{\beq {#1}\eeq}
\newcommand{\la}{\left\langle} \newcommand{\ra}{\right\rangle}

\newcommand{\A}{{\cal A}}
\newcommand{\s}{\sigma}
\newcommand{\f}{F_\A}
\def\lrf#1#2{ \left(\frac{#1}{#2}\right)}
\def\lrfp#1#2#3{ \left(\frac{#1}{#2}\right)^{#3}}

%

\renewcommand{\thefootnote}{\alph{footnote}}

\renewcommand{\thefootnote}{\fnsymbol{footnote}}
\preprint{DESY 06-071}
\title{Non-thermal Production of Dark Matter \\
  from Late-Decaying Scalar Field at Intermediate Scale}
\renewcommand{\thefootnote}{\alph{footnote}}

\author{Motoi Endo and Fuminobu Takahashi}

\affiliation{
  Deutsches Elektronen Synchrotron DESY, Notkestrasse 85,
  22607 Hamburg, Germany\\
  Institute for Cosmic Ray Research,
  University of Tokyo, Chiba 277-8582, Japan
}

\begin{abstract}
\noindent
We examine non-thermal dark matter production from a late-decaying
scalar field, with a particular attention on non-renormalizable
operators of $D=5$ through which the scalar field decays into the
standard model particles and their superpartners. We show that almost
the same number of superparticles as that of particles are generally
produced from the decay.  To avoid the gravitino overproduction
problem, the decay is favored to proceed via interactions with an 
intermediate cut-off scale $M \ll M_P$.  This should be contrasted 
to the conventional scenario using the modulus decay. The bosonic
supersymmetry partner of the axion, i.e., saxion, is proposed as a
natural candidate for such late-decaying scalar fields. We find that a
right amount of the wino/higgsino dark matter with a mass of $O(100)$
GeV is obtained for the saxion mass around the weak scale and axion
decay constant, $F_\A = O(10^{9-12})$ GeV.
\end{abstract}

\maketitle

\section{Introduction}
\label{sec:1}

The lightest superparticle (LSP) is one of the most plausible 
candidates for the dark matter (DM) in supersymmetric (SUSY) 
models with $R$-parity conservation. In the SUSY standard model, 
the lightest neutralino, composed of superpartners of the gauge 
and Higgs bosons,  is usually the LSP. 
From both the theoretical and experimental/observational points of view, the 
wino/higgsino DM is interesting, and it has been studied extensively 
so far~\cite{AM,Mirage,recent}. In this paper, we will concentrate 
on scenarios of the wino/higgsino DM production.

The present DM abundance is
\begin{eqnarray}
  \label{eq:dm}
  \Omega_{\rm DM} h^2 \;=\; 0.11 \pm 0.01,
\end{eqnarray}
from the latest WMAP three year data~\cite{Spergel:2006hy}, where $h$ 
is the present Hubble parameter in units of 100km/s/Mpc. The relic 
abundance of the LSP must fall in this range to account for the DM 
abundance. The relic LSP abundance depends on the thermal history 
of the universe. Once the universe is reheated up to high temperature 
after inflation, the LSP is thermally produced by particle scatterings 
and reaches thermal equilibrium. However, it is known that the thermal 
processes cannot provide a sufficient amount of the wino/higgsino LSP 
with a mass, $m_\chi = O(100)$ GeV, because of the large annihilation rates. 
Therefore, non-thermal production must take place below the decoupling 
temperature, $T_\chi \sim m_\chi/20$, to realize 
the observed DM abundance.

The non-thermal DM production at such a low temperature may be
realized by a late-time decay of a scalar field, $\phi$, with a mass
$m_\phi$. Renormalizable interactions generically induce a too rapid
decay into the standard model (SM) particles and their superpartners
unless the couplings are suppressed by some symmetries or mechanisms. 
Thus the decay should proceed via non-renormalizable operators suppressed 
by a large cut-off scale. In this paper, we focus on the $D=5$ operators. 
We will show that the branching ratio of the scalar decay into the SM
superparticles is comparable to that into the SM particles.

A modulus is one of such late-time decaying
fields~\cite{Moroi:1999zb}, because the interactions of the modulus
field are suppressed by the Planck scale, $M_P = 2.4 \times 10^{18}$
GeV.  The decay temperature of the modulus field is likely lower than
the decoupling temperature, leading to  the non-thermal production of
the wino/higgsino LSP.  However, according to the recent
works~\cite{Endo:2006zj,Kawasaki:2006gs, Asaka:2006bv,
Dine:2006ii,Endo:2006tf}, the late-time scalar decay is generically
plagued with the gravitino overproduction.  Thus the cosmological
scenario with a modulus field is strongly constrained, for example,
from the big--bang nucleosynthesis (BBN).

We point out that a natural source of the non-thermal DM production 
is related to {\it new physics at an intermediate scale}, $M \ll M_P$.
The gravitino overproduction problem is then relaxed. This is because
the total decay rate is enhanced as $\Gamma \sim m_\phi^3 / M^2$,
reducing the branching ratio of the gravitino production. Note also
that, since a light scalar field becomes acceptable, the decay into
the gravitinos may be kinematically forbidden.  In the following
sections we will show that a right amount of the wino/higgsino DM
is obtained by the scalar decay at an intermediate scale.  We propose
the saxion field, which is the bosonic SUSY partner of the axion, as a
natural candidate for such a scalar field.

Before closing the introduction, let us clarify differences from the
works in the past, which discussed the non-thermal DM production from
late-time decaying scalar fields. The decay from a heavy scalar with
Planck-suppressed couplings, i.e., a modulus field, has been extensively 
discussed in Ref.~\cite{Moroi:1999zb}. Also, Ref.~\cite{Khalil:2002mu} 
analyzed a similar subject with a lower cut-off, although they regarded 
the branching ratios of the superparticle as a free parameter. In this 
paper, we explicitly show that the branching ratio of the SM superparticle 
production is generically large, and that the effective number of the LSP 
produced by one scalar decay, $N_\chi$, is close to unity. This has a great 
impact on the non-thermal DM production scenarios.

The rest of the paper is organized as follows. In Sec.~\ref{sec:2} we
estimate the branching fraction of the SM (super)particle production
for the scalar decay through 
the non-renormalizable operators of $D=5$. We reexamine the
conventional scenario using the modulus decay for the non-thermal
DM production in Sec.~\ref{sec:3}. It is shown that the saxion decay
can account for the present DM abundance in Sec.~\ref{sec:4}. 
The last section is devoted to conclusion.

\section{Heavy Scalar Decay via $D=5$ Operators}
\label{sec:2}

In this section we consider a heavy scalar decay into the SM particles
and their superpartners, assuming the scalar mass $m_\phi$ much larger
than the weak scale.  Since those particles with R-parity odd
eventually produce at least one LSP per decay, the production rate of
the superparticles is crucial to determine the relic LSP abundance.
In the following, we assume that the scalar field is
R-parity even and singlet under the SM gauge groups. 
The purpose of this section is to estimate the decay rates of the
scalar field into the SM (super)particles via the non-renormalizable
operators of $D=5$. Although such operators have been studied in 
Ref.~\cite{Moroi:1999zb}, they overlooked some important interactions 
of the superparticle production. Below we investigate the operators 
one by one.

(I) Let us first consider the scalar decay into the gauge boson 
and gaugino. Such an interaction is effectively given by the dilatonic 
coupling with a cut-off scale $M$,
\begin{eqnarray}
  \mathcal{L} \;=\; \frac{\lambda_G}{M}
  \int d^2\theta\, \phi W^{(a)} W^{(a)},
\end{eqnarray}
where $\lambda_G$ is a numerical coefficient
and $W^{(a)}$ is the supersymmetric field strength of the SM gauge
supermultiplet. Including the supergravity effects, we obtain
\begin{eqnarray}
  \mathcal{L}_G &\simeq& 
  \frac{\lambda_G}{M}
  \bigg[ 
  \frac{\phi_R}{\sqrt{2}} 
  \left( -\frac{1}{4} F_{\mu\nu}^{(a)} F^{(a) \mu\nu} \right) +
  \frac{\phi_I}{\sqrt{2}} 
  \left( -\frac{1}{8} \epsilon^{\mu\nu\rho\sigma} 
    F_{\mu\nu}^{(a)} F_{\rho\sigma}^{(a)} \right) 
  \nonumber \\ && 
  \left.+\frac{1}{4} e^{G/2} \left\{
    \left( G^{\bar{\phi}}_{\phantom{a}\phi} \phi + 
      G^{\bar{\phi}}_{\phantom{a}\bar{\phi}} \phi^\dagger \right)
      \,\bar{\lambda}^{(a)} {\mathcal P}_R \lambda^{(a)}
      + 
      \left( G^{\phi}_{\phantom{a}\bar{\phi}} \phi^\dagger + 
        G^{\phi}_{\phantom{a}\phi} \phi \right)
      \,\bar{\lambda}^{(a)} {\mathcal P}_L \lambda^{(a)}
    \right\}
    \right],
\end{eqnarray}
where the real and imaginary components of the scalar field are
defined as $\phi \equiv (\phi_R + i\phi_I)/\sqrt{2}$. Although there
are the other contributions from the gaugino kinetic terms, the
resultant decay rate through them is proportional to the gaugino mass
squared, $m_\lambda^2$, after applying the equation of motion.
Therefore they are neglected unless $m_\phi$ is close to (but larger than)
$2 m_\lambda$. 
The subindex, $i$, attached to the total K\"ahler potential, $G = K 
+ \ln |W|^2$, represents a derivative with respect to the field, $i$, 
while the superscript is obtained by $G^i = g^{ij^*} G_{j^*}$, where 
$g^{ij^*}$ is the inverse of the K\"ahler metric $g_{ij^*}$. The coefficient 
$e^{G/2}$ is equal to the gravitino mass at the vacuum, $m_{3/2} = \langle 
e^{G/2} \rangle$.

The coefficients of the interactions with the gauginos are 
\begin{eqnarray}
  G^{\phi}_{\phantom{A}\bar{\phi}} &=& 
  (g^{\phi\bar\phi}G_{\bar\phi})_{\bar{\phi}} \;=\;
 g^{\phi\bar\phi} G_{\bar \phi\bar \phi} + 
 (g^{\phi\bar\phi})_{\bar \phi} G_{\bar\phi},
  \\
  G^{\phi}_{\phantom{a}\phi} &=& 
  (g^{\phi\bar\phi}G_{\bar\phi})_{\phi} \;=\;
  1 +  (g^{\phi\bar\phi})_{\phi} G_{\bar\phi},
\end{eqnarray}
where we have assumed $g_{ij^*} = 0$ for $i \ne j$, for simplicity.
If $m_\phi \lsim m_{3/2}$, we can see
$G^{\phi}_{\phantom{A}\bar{\phi}} + G^{\phi}_{\phantom{a}\phi} = O(1)$
barring cancellations. In fact, the minimal K\"ahler potential without
the SUSY mass term (i.e., $G_{\phi \phi }= 0$) gives
$\,G^{\phi}_{\phantom{A}\bar{\phi}} + G^{\phi}_{\phantom{a}\phi} =
1$. On the other hand, if the mass is enhanced by the SUSY mass,
namely $m_\phi \simeq e^{G/2} |G_{\phi\phi}| \gg m_{3/2}$, the
coefficient is enhanced as $G^{\phi}_{\phantom{A}\bar{\phi}} +
G^{\phi}_{\phantom{a}\phi} = O(m_\phi/m_{3/2})$.  However if the
scalar mass $m_\phi$ is enhanced due to the non-SUSY effects, e.g.,
$\delta K =|\phi|^2 |z|^2/M^2$ with the SUSY breaking field $z$, the
coupling is given by $G^{\phi}_{\phantom{A}\bar{\phi}} +
G^{\phi}_{\phantom{a}\phi} = O(1)$.

Then the decay rates into the gauge multiplets are evaluated 
as~\cite{Endo:2006zj}
\begin{eqnarray}
  \Gamma_G \;\equiv\;
  \Gamma(\phi \rightarrow gauge\ boson) \;\simeq\;
  \Gamma(\phi \rightarrow gaugino) \;\simeq\;
  N_g \frac{|\lambda_G|^2}{8\pi} \frac{m_\phi^3}{M^2},
 \label{eq:gauge-decay}
\end{eqnarray}
either if $m_\phi \sim m_{3/2}$ or if $m_\phi \gg m_{3/2}$ due to the
SUSY mass term.  Here $N_g$ is the number of the possible final
states, and $N_g = 12$ for the SM gauge groups, $SU(3)_C \times
SU(2)_L \times U(1)_Y$. We find that the production rate of the
gaugino is comparable to that of the gauge boson, and is {\it not}
suppressed by the gaugino mass. On the other hand, if the scalar mass
is smaller than $m_{3/2}$, the gaugino production rate is estimated as
$\Gamma \sim m_{3/2}^2 m_\phi/M^2$ up to a numerical factor, while the
gauge boson production rate is the same as
(\ref{eq:gauge-decay}). Similarly, if $m_{\phi}$ is enhanced due to
the non-SUSY mass term as $m_\phi \gg m_{3/2}$, the gaugino production
rate is $\Gamma \sim m_{3/2}^2 m_\phi/M^2$, which is suppressed
compared to the gauge boson production rate $\Gamma \sim
m_\phi^3/M^2$.

(II) The scalar field can couple with the matter fields in the
K\"ahler potential. When the K\"ahler potential includes such a
non-renormalizable term as, $K = \phi Q Q^{'\dagger}/M + h.c.$, in
which the matter fields $Q$ and $Q'$ appear with the opposite 
chiralities, we can show that the decay rate into the SM 
(super)particle is proportional to powers of the masses 
of the matter fields.

Let us first consider the matter scalar production rate. The first
contribution comes from the kinetic term of the scalar fields:
$\mathcal{L}_K = g_{ij^*} \partial_\mu \phi^i \partial^\mu \phi^{j*}$. 
Then the decay rate is proportional to the fourth power of the mass 
of the matter scalar after using the equation of motion, since the 
interaction term becomes $\mathcal{L}_K = -(\phi/M)\,\tilde Q 
(\partial^2 \tilde Q')$ up to a total derivative term, where 
$\tilde Q$ is the scalar component of the matter chiral multiplet 
$Q$~\cite{Moroi:1999zb}. Thus the decay rate is estimated as 
$\Gamma_K \sim (m_{Q'}/m_\phi)^4 \times m_\phi^3/M^2$.

The other couplings arise from the scalar potential. In particular,
the decay rate may be enhanced due to the supergravity effects when 
$m_{3/2}$ is large. Since non-renormalizable operators violate the 
conformal symmetry, the operator receives corrections of order $m_{3/2}^2$ 
at the leading level. Thus we obtain $\mathcal{L}_V \sim \frac{1}{M} 
m_{3/2}^2 \phi \tilde Q \tilde Q^{'\dagger} + h.c.$, though the coefficient 
depends on details of the models. The decay rate then becomes $\Gamma_V 
\sim (m_{3/2}/m_\phi)^4 \times m_\phi^3/M^2$.

The above supergravity effects seem to give a correction to the matter
scalar mass, $\delta m_Q^2 \sim m_{3/2}^2$, by taking a vacuum
expectation value (VEV), $\la\phi\ra \sim M$. It should be noted,
however, that this estimate is too naive and actually overestimates
the correction when $m_{3/2}$ is larger than the weak scale.  This is
because there exists a cancellation among the contributions to the
soft scalar mass if $|G_\phi| \ll \la \phi \ra$. This cancellation can
be understood in terms of the K\"ahler invariance. To be concrete,
let us consider the minimal K\"ahler potential with the coupling
$\delta K = \phi Q Q^{'\dagger}/M + h.c.$. Since the above contribution 
in $O(m_{3/2}^2)$ arises from $K_\phi W$ of the scalar potential, it 
is always accompanied by $W_\phi$.  Thus when the F-term of $\phi$ is
suppressed, the correction of $O(m_{3/2}^2)$ will cancel with that
from $W_\phi$ after taking the VEV.

Lastly, the production rates of the matter fermions, $\psi_Q$, are
chirally suppressed. The decay rate is estimated as $\Gamma_{\psi_Q}
\sim (m_{\psi_Q}/m_\phi)^2 \times m_\phi^3/M^2$, which is proportional
to the fermion mass squared.

(III) The heavy scalar field can decay into the Higgs supermultiplets
via the interaction:
\begin{eqnarray}
  \mathcal{L} \;=\; \frac{\lambda_H}{M} 
  \int d^4\theta\, \phi^\dagger H_u H_d + h.c.,
  \label{eq:higgs}
\end{eqnarray}
though such a coupling is related to the $\mu$-problem and it is quite
model-dependent whether or not such a coupling exists.  We notice that
the Higgs chiral supermultiplets appear with the same chirality.
Expanding the interaction (\ref{eq:higgs}), we obtain
\begin{eqnarray}
  \mathcal{L}_H &\simeq& \frac{\lambda_H}{M} 
  \left[ -(\partial^2 \phi^\dagger) H_u H_d + 
    e^{G/2} G^{\bar \phi}_{\phantom{\phi}\phi} 
    \phi \bar{\tilde H}_{uR} \tilde H_{dL} 
  \right]
  + {\rm h.c.}.
\end{eqnarray}
In addition, the scalar trilinear coupling $\phi^\dagger H_u H_d$ arises 
from the scalar potential; the decay rate through the supergravity corrections 
is given by $\sim |\lambda_H|^2 m_{3/2}^2 m_\phi/M^2$.  We neglect 
the interaction terms whose  coupling  is suppressed by the
higgsino mass. In particular, 
they include the $U(1)$ connection term in the covariant derivative 
of the higgsino kinetic term, and that obtained by expanding $e^{G/2}$ 
in  the higgsino mass term. These are generically 
small in the mass eigenstate basis by considering the interference with the 
SUSY breaking field~\cite{Dine:2006ii,Endo:2006tf}.

Thus, either if $m_\phi \sim m_{3/2}$ or if $m_\phi \gg m_{3/2}$ due to
the SUSY mass term, the decay rates are given by
\begin{eqnarray}
  \Gamma(\phi \rightarrow HH) \;\simeq\; 
  \Gamma(\phi \rightarrow \tilde H\tilde H) \;\simeq\;   
  \frac{|\lambda_H|^2}{8\pi} \frac{m_\phi^3}{M^2}
  \label{eq:higgsdecay}
\end{eqnarray}
for the Higgs multiplets which are doublet under the $SU(2)_L$
symmetry.  We find that the decay rates are comparable to $\Gamma_G$
if $\lambda_G \sim \lambda_H$, and that the higgsino production rate
is the same order of the Higgs production rate. In contrast, when the
scalar mass is enhanced by the non-SUSY mass term as $m_\phi \gg
m_{3/2}$, the higgsino production is suppressed compared to the decay
into the Higgs bosons by $O(m_{3/2}^2/m_\phi^2)$.

(IV)The following interactions in the superpotential permit the
three-body decay,
\begin{eqnarray}
  \mathcal{L} \;=\; 
  \frac{\lambda_Y}{M} \int d^2\theta\, \phi H_u T^c Q + h.c.,
\end{eqnarray}
where $H_u, T, Q$ are the chiral supermultiplet of the up-type Higgs,
right-handed top quark and left-handed quark doublet of the third
generation, respectively. Then the interactions become
\begin{eqnarray}
  \mathcal{L}_Y \;=\; 
  \frac{\lambda_Y}{M} \left[
    \phi H_u \bar t q + 
    \phi \tilde t^* \bar{\tilde H}_{uR} q + 
    \phi \bar t \tilde H_{uL} \tilde q 
  \right] + {\rm h.c.}, 
\end{eqnarray}
in the global SUSY Lagrangian. Although the decay rates do not receive
chirality suppressions, the decay is a three-body process and the
phase space is small~\cite{Moroi:1999zb}:
\begin{eqnarray}
  \Gamma(\phi \rightarrow H t q) \;\simeq\;
  \Gamma(\phi \rightarrow \tilde H \tilde t q) \;\simeq\;
  \Gamma(\phi \rightarrow \tilde H t \tilde q) \;\simeq\;
  \frac{|\lambda_Y|^2}{256\pi^3} \frac{m_\phi^3}{M^2},
  \label{three}
\end{eqnarray}
for $m_\phi \gg m_{\rm soft}$. Note that couplings with the lighter
quarks and leptons must be suppressed, since the corresponding Yukawa
coupling constants need fine-tunings by taking account of the scalar 
VEV, $\la\phi\ra \sim M$.

Additionally, the scalar couplings receive supergravity corrections.
The decay rate through the correction which is of order $m_{3/2}$ 
for the operator $\phi H_u \tilde u^* \tilde q$ is suppressed both by the 
phase space and by the gravitino mass squared. 
If $m_\phi < m_{3/2}$, the decay rate becomes larger than (\ref{three}).
At first sight, the scalar 
VEV seems to induce a correction of an order $m_{3/2}$ to the trilinear 
scalar coupling, $H_u \tilde u^* \tilde q$. However, it is actually
suppressed if $|G_\phi| \ll \la \phi \ra$, since the scalar $\phi$ 
contributes to the A-parameter through the auxiliary component $G_\phi$ 
due to the K\"ahler invariance.

\hspace{1cm}

From the analyses of (I) -- (IV), either if  $m_\phi \sim m_{3/2}$ or if
$m_\phi \gg m_{3/2}$ due to the SUSY mass $G_{\phi\phi} \gg 1$,
almost the same amount of the SM superparticles is produced as that of the 
SM particles:
\begin{eqnarray}
  {\rm Br}(\phi \rightarrow {\rm SM\ particles}) 
  \;\simeq\; 
  {\rm Br}(\phi \rightarrow {\rm SM\  superparticles}),
\end{eqnarray}
with the total decay rate
\begin{eqnarray}
  \Gamma_\phi \;=\; \frac{c}{4\pi} \frac{m_\phi^3}{M^2},
  \label{eq:scalar_decayrate}
\end{eqnarray}
where $c$ is a numerical coefficient, and given by $c \simeq
N_g\lambda_G^2 + \lambda_H^2$, for the dilatonic coupling and the
interaction with the Higgs supermultiplets. The other decay processes
are suppressed either by the soft masses or by the phase space. On the
other hand, when the scalar mass is enhanced by the non-SUSY mass term
as $m_\phi \gg m_{3/2}$, the above superparticle production rate becomes 
suppressed compared to that of the SM particle. Note that the single-gravitino 
production induce a serious problem in this case, 
as will be discussed in the next 
section. If $m_{3/2} > m_\phi$, the supergravity corrections significantly 
affect the decay rates, and the branching ratio of the superparticle 
production depends on the non-renormalizable operators. We emphasize 
again that the branching ratio of the superparticle production is 
generically sizable, ${\rm Br}(\phi \to {\rm SM\ superparticles}) = 
O(0.1)$, either when $m_\phi \sim m_{3/2}$ or if $m_\phi \gg m_{3/2}$ due 
to the SUSY mass term.

\section{Difficulties in Modulus Decay}
\label{sec:3}

It has been known that the wino/higgsino LSP with a mass of 
$O(100)$ GeV cannot be produced sufficiently in the thermal 
bath because of its large annihilation cross section. We need 
therefore the non-thermal process below the freeze-out temperature 
in order to explain the observed DM abundance. In this section let 
us reexamine the non-thermal DM production from the modulus decay.

The modulus field corresponds to a dynamical degree of freedom
associated with a flat direction of the scalar potential. Since 
the SUSY breaking effects stabilize the potential, the modulus 
field naturally obtains a mass around the gravitino mass, $m_{3/2}$. 
In addition, the modulus mass can be even larger with sizable
non-perturbative corrections. The modulus decay is generally 
induced by the non-renormalizable operators, whose couplings 
are suppressed by the Planck scale, namely $M = M_P$ in the 
previous section. Such a light and late-decaying scalar field 
is likely to dominate the energy of the universe before its decay. 
Then the subsequent thermal history depends on the decay processes 
of the modulus field, that is, the decay temperature and the branching 
fractions of the decay products. The modulus mass is required to be 
$m_\phi \gsim 100$ TeV~\cite{Moroi:1999zb}~\footnote{ 
  One might consider the modulus decay via the $D=6$ operators. 
  However it is unlikely that the decay takes place before the 
  BBN starts.
}, from the cosmological BBN bound.

If the decay takes place after the freeze-out of the LSP, the relic
LSP abundance is estimated in terms of the following parameters: the
annihilation cross section of the LSP, $\la \sigma v \ra$, the modulus
decay temperature, $T_d$, and the branching fraction of the
superparticle production, which provides an effective number of the
LSP from one modulus decay, $N_\chi$.  According to the previous section, 
the branching fraction of the LSP production is generically quite
large. Then the resultant LSP abundance after the pair annihilation
becomes independent of the initial abundance, and given
by~\cite{Fujii:2001xp}
\begin{eqnarray}
  Y_{\rm LSP} \;\simeq\; 
  \sqrt{\frac{45}{8 \pi^2 g_{*d}}} \frac{1}{\la \sigma v \ra M_P T_d },
\end{eqnarray}
or equivalently,  the relic LSP density is
\begin{eqnarray}
 \Omega_\chi h^2 \;\simeq\;
  0.5
  \lrfp{10.75}{g_{*d}}{\frac{1}{2}} 
  \lrf{10^{-7}\ {\rm GeV}^{-2}}{\la \sigma v \ra} 
  \lrf{m_\chi}{100\ {\rm GeV}} 
  \lrf{100\ {\rm MeV}}{T_d},
\end{eqnarray}
where $m_\chi$ is the LSP mass.  In the case of the wino LSP, the
annihilation cross section is estimated to be e.g. $\langle \sigma v
\rangle \simeq 3.3 \times 10^{-7}$ GeV$^{-2}$ for the wino mass $M_2 =
100$ GeV, and that of the higgsino LSP is roughly the same order of 
the magnitude (see Eqs.~(\ref{eq:ann_for_wino}) and 
(\ref{eq:ann_for_higgsino})).  Since the annihilation cross section
decreases for the larger LSP mass, it is likely smaller than
$O(10^{-7})\,$GeV$^{-2}$.  Thus the decay temperature must satisfy
$T_d \gsim 100$ MeV in order to obtain a right amount of the relic 
DM abundance (\ref{eq:dm}).

Using the total decay rate (\ref{eq:scalar_decayrate}) with $M = M_P$, 
the decay temperature is given by
\begin{eqnarray}
  T_d \;\simeq\; 5.5~\mathrm{MeV} \cdot
  c^\frac{1}{2}\left({m_\phi \over 100~\mathrm{TeV}}\right)^{3/2}\;,
\end{eqnarray}
where $c = O(1)$. To satisfy $T_d \;\gsim\; 100$ MeV, the modulus mass 
should be large enough:
\begin{eqnarray}
  m_\phi \;\gsim\; 10^{6-7}\ {\rm GeV}.
\end{eqnarray}
The bound on the modulus mass becomes severer for smaller $c$ and the
heavier wino/higgsino. It should be stressed again that one of the
crucial points to derive the above constraint is an effective number
of the LSP from the modulus decay. As long as the decay proceeds
through the non-renormalizable operators of $D=5$, the branching ratio
of the LSP production is generically large, and close to unity. This 
is another difficulty of the modulus decay, and let us call it as 
``the moduli-induced LSP problem''.

The mass of the modulus field depends on the mechanism of the
potential stabilization. A modulus field as heavy as $m_\phi \gsim
10^{6-7}$ GeV may be realized in the scenarios of the mixed
modulus-anomaly/KKLT mediation~\cite{Mirage} and in the racetrack
setups~\cite{racetrack}.  However, such a heavy modulus generically
suffers from the gravitino overproduction~\cite{Endo:2006zj}. 
Since the conformal anomaly mediates the SUSY breaking effects 
to the visible sector and the correction is related to the gravitino 
mass~\cite{AM}, the gravitino mass is favored to be $\lsim 100$ TeV, 
leading to the relation $m_\phi > 2 m_{3/2}$. The production rate of the 
goldstino, which is a longitudinal component of the gravitino, can be 
enhanced in the mode $\phi \to 2 \psi_{3/2}$ in high energy limit $p_\mu 
\gg m_{3/2}$. Actually, the decay rate is obtained as~\cite{Endo:2006zj}
\begin{eqnarray}
  \Gamma(\tilde\Phi \rightarrow 2\psi_{3/2}) \;\simeq\;
  \frac{|{\cal G}^{\rm (eff)}_\Phi|^2}{288\pi} 
  \frac{m_\phi^5}{m_{3/2}^2 M_P^2},
\end{eqnarray}
where $\tilde{\Phi}$ denotes the mass eigenstate (mainly composed of
$\phi$), and ${\cal G}^{\rm (eff)}_\Phi$ is the effective auxiliary
field of $\tilde \Phi$.  The auxiliary component of the scalar field
crucially depends on the structure of the SUSY breaking sector as was
pointed out in Ref.~\cite{Dine:2006ii}. Since both the modulus and
SUSY breaking fields are not generally protected by any symmetries at
the vacuum, they couple with each other, and the rate becomes
sizable~\footnote{
  Even if the modulus does not couple to the SUSY breaking sector,
  the branching fraction of the gravitino production is still large.
  If the scalar mass of the SUSY breaking field, $m_z$, is smaller 
  than $m_\phi$ as $B_{3/2} = O(10^{-4}-10^{-3})$, while 
  $B_{3/2} = O(0.01-0.1)$ if $m_z > m_\phi$.
  Such large branching ratio causes serious cosmological 
  problem~\cite{Kohri:2004qu}.
}. In fact, the effective auxiliary field can be as large as
${\cal G}^{\rm (eff)}_\Phi = \kappa_{3/2} m_{3/2}/m_\phi$ with
$\kappa_{3/2} = O(1)$~\cite{Endo:2006tf}. Thus the branching ratio of
the gravitino production is $B_{3/2} = O(0.01-0.1)$. Such models with
a large gravitino production is severely constrained by the cosmology, 
which we call ``the moduli-induced gravitino problem''~\cite{Endo:2006zj}.

As a final remark of this section, let us mention the case of a modulus 
mass coming from the non-SUSY mass which satisfies $m_\phi \gg m_{3/2}$. 
In this case, a mass of the fermionic component of $\phi$ is likely much 
smaller than $m_\phi$, so the single-gravitino production starts to dominate 
over the pair production. The decay rate is~\cite{Buchmuller:2004rq}
\begin{eqnarray}
  \Gamma(\phi \rightarrow \psi_\phi \psi_{3/2})
  \;\simeq\;
  \frac{1}{48\pi} \frac{m_\phi^5}{m_{3/2}^2M_P^2}
\end{eqnarray}
for $m_\phi \gg m_{\psi_\phi}, m_{3/2}$, which dominates over the other
decay processes. It is noticed that the rate is independent of the 
VEV of the effective auxiliary field ${\cal G}^{\rm (eff)}_\Phi$. Thus 
such modulus field generally suffers from the production of too much 
gravitinos.

In summary, the modulus decay is disfavored as a source of the 
non-thermal DM production because of the above ``the moduli-induced 
gravitino/LSP problems.'' In the next section, we propose a natural 
solution which is free from these problems and also explains the 
relic wino/higgsino DM abundance.

\section{Dark Matter from Saxion Decay}
\label{sec:4}
The difficulties of the modulus decay in producing a right amount of
DM are associated with the fact that the interactions are suppressed
by the Planck scale, as shown in the previous section. The modulus
mass cannot be smaller than $100$~TeV to decay before BBN, which makes
the modulus decay into the gravitinos almost inevitable. One of the
attractive ways to get around the problem is to postulate a cut-off
scale smaller than the Planck scale; the branching ratio of the
gravitino production can be then suppressed, or may even be
kinematically forbidden.  Thus an intermediate-scale physics is likely
to play an important role in the non-thermal wino/higgsino DM
production.

An intermediate-scale physics naturally arises in the Pecci-Quinn (PQ)
mechanism~\cite{Peccei:1977hh,Wilczek:1977pj,Kim:1979if,Dine:1981rt}.
The strong CP problem is one of the profound puzzles in SM, 
and the PQ mechanism using a pseudo Nambu-Goldstone boson
called axion provides an elegant solution to the problem.  The axion
field has a flat potential at a perturbative level, and acquires a
finite mass from non-perturbative QCD effects via the anomaly. The
axion rolls down to the potential minimum at which the strong CP phase
vanishes. From the cosmological/astrophysical observations, the axion
decay constant, $F_\A$, is required to satisfy, $F_\A \gsim 10^9$ GeV
from SN1987a~\cite{Raffelt:1996wa,Kolb:1990vq}, and $F_\A \lsim
10^{12-15}$ GeV by the axion-overclosure limit~\footnote{
  As is well known, the axion itself is a DM candidate.
  However the contribution depends on the initial condition 
  of the axion field and the subsequent thermal history. In 
  the following, we will focus  on the abundance of the 
  wino/higgsino LSP.
}, although the latter
depends on the cosmological
scenario~\cite{Preskill:1982cy,Turner:1985si,Kawasaki:1995vt}.

In the SUSY theories, the axion field is extended to the axion
supermultiplet, $\A$. The multiplet contains two bosonic real fields,
which are axion and saxion, and one chiral fermion called axino. 
The scalar component of the axion multiplet is represented as
\begin{eqnarray}
  \A \;\equiv\; \frac{1}{\sqrt{2}} (\sigma + ia),  
\end{eqnarray}
where $\s$ and $a$ are saxion and axion fields, respectively.  The
saxion behaves like a modulus in the evolution of the universe. Since
the saxion potential as well is flat at a perturbative level in the
SUSY limit, the field is stabilized by the SUSY breaking (and perhaps
non-perturbative corrections), and it acquires a mass around the
gravitino mass, $m_\sigma \sim m_{3/2}$~\footnote{
  Precisely speaking, the saxion mass is model-dependent. For instance, in 
  the heavy gravitino setups, if the axion multiplet is sequestered from 
  the SUSY breaking sector, the saxion mass can be smaller than the gravitino 
  mass by a loop factor.
}. 
The initial position of the saxion generically deviates from the potential 
minimum in the early universe, and it starts to oscillate when the Hubble 
parameter becomes comparable to the saxion mass. Then the energy fraction 
of the saxion can be sizable, and large number of the superparticles may 
be produced at the decay~\footnote{
  The LSP may be the fermionic superpartner of the axion, axino. However the axino
  mass depends on the models, and we assume that the wino/higgsino is the 
  LSP in the following. 
}.
 
It is a strength of the couplings with the SM sector that significantly 
differs from the modulus field. The axion supermultiplet interacts with 
the SM (super)particles with the couplings suppressed by the axion decay 
constant, $F_\A$. Since $F_\A$ must be much smaller than the Planck scale 
because of the cosmological requirement, the saxion decay rate is much 
larger than that of the modulus field with the same mass. In particular, 
the axion multiplet $\A$ necessarily has a dilatonic coupling:
\begin{eqnarray}
  \label{eq:aWW}
  {\mathcal L} \;=\; -\int d^2\theta\, 
  \frac{\alpha}{4\sqrt{2} \pi} 
  \frac{{\mathcal A}}{F_\A} W^{(a)} W^{(a)},
  \label{eq:axion_gauge}
\end{eqnarray}
where $\alpha$ is a gauge coupling constant, and $\f$ the axion 
decay constant. Here we adopt the hadronic axion model~\cite{Kim:1979if} 
for simplicity, and a brief comment on the case of the DFSZ axion 
model~\cite{Dine:1981rt} will be given later. The dilatonic coupling 
arises from the chiral anomaly of the PQ symmetry which is extended 
to the complex $U(1)_{PQ}$ symmetry in the SUSY framework. The interaction 
(\ref{eq:axion_gauge}) actually reproduces the axion coupling with the 
$SU(3)_C$ gauge bosons, providing a solution to the strong CP problem. 
It also provides the saxion couplings with the gauge bosons and gauginos:
\begin{eqnarray}
  {\mathcal L} &=& 
  \frac{\alpha}{8\pi}\frac{\sigma}{F_\A} F_{\mu\nu}^{(a)} F^{(a)\mu\nu} 
  - \frac{\alpha}{8\pi}\frac{\sigma}{F_\A} e^{G/2}
  \bigg[ 
  \left(G^{A}_{\phantom{A}A} + G^{A}_{\phantom{A}\bar A} \right)
  \bar\lambda^{(a)} {\mathcal P}_L \lambda^{(a)} + {\rm h.c.}
  \bigg],
\end{eqnarray}
where $F_{\mu\nu}^{(a)}$ and $\lambda^{(a)}$ denote the field strength
of the gauge boson and the gaugino field, respectively. The decay
rates of the saxion into the gauge bosons $g$ and gauginos $\lambda$
are given by~\footnote{ 
  The saxion may have a coupling with the axion, which enables the saxion 
  to decay into two axions. If the saxion dominantly decays into the axions, 
  it can be cosmologically problematic~\cite{Hashimoto:1998ua}. However, it 
  is model-dependent whether this process dominates the saxion decay. In fact, 
  such a coupling can be suppressed. In the following, we simply assume that
  the decay mode is suppressed. 
}
\begin{eqnarray}
  \label{eq:saxion_decayrate}
  \Gamma (\sigma \rightarrow gg) \;\simeq\; 
  \frac{3\alpha^2}{64\pi^3} 
  \frac{m_\sigma^3}{F_\A^2},
  \quad\quad
  \Gamma (\sigma \rightarrow \lambda\lambda) \;\simeq\; 
  \frac{3\alpha^2}{64\pi^3}  |\kappa|^2 \frac{m_\sigma^3}{F_\A^2}
\end{eqnarray}
with
\begin{eqnarray}
  \kappa \;\equiv\; 
  \left(G^{A}_{\phantom{A}A} + G^{A}_{\phantom{A}\bar A}\right)
  \frac{m_{3/2}}{m_\sigma},
\end{eqnarray}
where we have neglected the sub-leading terms suppressed by
$(m_\lambda/m_\sigma)^2$ ($m_\lambda$: the gaugino mass).  Although we
have assumed that the saxion universally couples to the  $SU(3)_C \times
SU(2)_L \times U(1)_Y$ gauge multiplets in estimating the decay rates,
this assumption is not crucial for the following arguments. This is
because the saxion decay into the gauge bosons is dominated by that
into gluons, to which the QCD axion multiplet necessarily couples.
Thus the branching ratio of the gaugino production is ${\rm Br}(\sigma
\rightarrow \lambda \lambda) \simeq |\kappa|^2/(1 + |\kappa|^2)$,
which depends on $\kappa$.  As discussed in Sec.~\ref{sec:2},
$G^{A}_{\phantom{A}A} + G^{A}_{\phantom{A}\bar A}$ is generically
$O(1)$ for $m_\s \lesssim m_{3/2}$, leading to $\kappa = O(1)$. 
In fact, for the minimal K\"ahler potential we have
$(G^{A}_{\phantom{A}A} + G^{A}_{\phantom{A}\bar A}) = 1$. Here it
should be note that the axion multiplet is not allowed to have nonzero
SUSY mass, since the potential minimum of the axion would be shifted
otherwise.  On the other hand, if a mass hierarchy, $m_\sigma \gg
m_{3/2}$, is realized, $\kappa$ can be suppressed as
$O(m_{3/2}/m_\s)$.  Thus the branching ratio of the gaugino production
is sizable in the saxion decay,
\begin{eqnarray}
  \label{eq:saxion_gauginoBr}
  {\rm Br}(\sigma \rightarrow \lambda \lambda) \;=\;
  \frac{|\kappa|^2}{1 + |\kappa|^2} = O(0.1),
\end{eqnarray}
unless the saxion mass is much larger than $m_{3/2}$ due to the
non-SUSY mass.  It should be stressed that the branching ratio of the
gaugino production is not suppressed by the gaugino mass as noted in
the previous section, and that the decay rates are enhanced by
$O(M_P^2/F_\A^2)$ compared to the modulus case.

The relic wino/higgsino LSP abundance is determined by solving the
Boltzmann equations. The wino/higgsino LSP can be produced from
thermal scatterings, but it is known that the thermal relic abundance
is too small to account for the present DM density.  As will be shown
below, the saxion decay can non-thermally produce a large enough
number of LSPs, due to both large branching ratio of the gaugino
production and relatively low decay temperature.

The LSP number density $n_\chi$ evolves as
\begin{eqnarray}
\label{eq:boltz}
  \dot{n_\chi}+3 H n_\chi \;=\; - \la \sigma v \ra n_\chi^2, 
\end{eqnarray}
where we have neglected the production of the LSPs from thermal
scattering processes by assuming that the saxion decay temperature is
lower than the decoupling temperature of the LSP, $T_\chi \sim
m_\chi/20$.  Here $\la \sigma v \ra$ is the thermally averaged
annihilation cross section of the LSP.  Let us rewrite
Eq.~(\ref{eq:boltz}) in terms of the LSP abundance, $Y_\chi \equiv
n_\chi/s$,
\begin{eqnarray}
  \frac{d Y_\chi}{d T} \;=\; \sqrt{\frac{8 \pi^2 g_{*d}}{45}} 
  \la \sigma v \ra M_P Y_\chi^2,
\end{eqnarray}
where $g_{*d}$ is an effective number of massless degrees of freedom
at the saxion decay, and we have approximated that $g_{*d}$ is almost
constant during the evolution. By integrating this equation, we obtain
the simple analytic formula of the relic abundance for $T <
T_d$~\cite{Fujii:2001xp}:
\begin{eqnarray}
\label{eq:y-chi}
  Y_\chi(T) \;=\; \left[Y_\chi(T_d)^{-1} + \sqrt{\frac{8 \pi^2 g_{*d}}{45}} 
    \la \sigma v \ra M_P (T_d - T)\right]^{-1}.
\end{eqnarray}
If  the initial LSP abundance $Y_\chi(T_d)$  satisfies 
\bea
Y_\chi(T_d) &\gg & 
Y^{(c)}_\chi \equiv \left[  \sqrt{\frac{8 \pi^2 g_{*d}}{45}} 
    \la \sigma v \ra M_P T_d      \right]^{-1},\non\\
&\simeq &8.4 \times 10^{-12} \xi^{-1} 
  \lrfp{10.75}{g_{*d}}{\frac{1}{4}} 
  \lrf{10^{-7}\ {\rm GeV}^{-2}}{\la \sigma v \ra}
  \lrfp{1\ {\rm TeV}}{m_\s}{\frac{3}{2}}
  \lrf{\f}{10^{12}\ {\rm GeV}},  
  \label{eq:ineqY}
\eea
 the second term in the parenthesis of Eq.~(\ref{eq:y-chi}) dominates
 over the first term. Then the final abundance is approximately given
 by
\begin{eqnarray}
  \label{eq:relic_abundace}
  Y_\chi^{(f)} \;\simeq\; 
  \sqrt{\frac{45}{8 \pi^2 g_{*d}}} \frac{1}{\la \sigma v \ra M_P T_d }.
\end{eqnarray}
Therefore the relic abundance is determined only by the annihilation
cross section of the LSP and the decay temperature of the parent
particle, independent of the initial amount of the LSP.

The annihilation cross section, $\la \sigma v \ra$, for the wino LSP 
is evaluated as~\cite{Moroi:1999zb}
\begin{eqnarray}
  \label{eq:ann_for_wino}
  \la \sigma v \ra \;\simeq\; 
  \frac{g_2^4}{2 \pi} m_\chi^{-2} \frac{(1-x_W)^{3/2}}{(2-x_W)^2},  
\end{eqnarray}
where $x_W \equiv m_W^2/m_\chi^2$ denotes the mass squared ratio of
the W boson and the LSP(wino), and $g_2$ is the gauge coupling constant of
$SU(2)_L$. For $m_\chi = O(10^2)$ GeV, $\la \sigma v \ra$ is roughly
estimated as $O(10^{-7})$ GeV$^{-2}$.  For the higgsino LSP, it is
given by~\cite{Olive:1989jg}
\begin{eqnarray}
  \label{eq:ann_for_higgsino}
  \la \sigma v \ra \;\simeq\; 
  \frac{g_2^4}{32 \pi} m_\chi^{-2} \frac{(1-x_W)^{3/2}}{(2-x_W)^2},    
\end{eqnarray}
which is smaller than the wino cross section by one order of magnitude.

Assuming that the saxion dominantly decays into the gauge bosons 
and gauginos with the rates given by Eq.~(\ref{eq:saxion_decayrate}), 
the decay temperature is given by
\begin{eqnarray}
  \label{eq:td}
  T_d \;\simeq\; 1.1\times 10^2\,{\rm MeV} \,\xi\,
  \lrfp{10.75}{g_{*d}}{\frac{1}{4}} 
  \lrfp{m_\s}{1\,{\rm TeV}}{\frac{3}{2}}
  \lrf{10^{12}\,{\rm GeV}}{F_\A}, 
\end{eqnarray}
where $\xi$ is defined as $\xi \equiv \sqrt{(1+|\kappa|^2)/2}$. 
Using Eqs.~(\ref{eq:relic_abundace}) and (\ref{eq:td}), 
the LSP relic density is given by
\begin{eqnarray}
  \label{eq:omega_LSP}
  \Omega_\chi h^2 \simeq 0.23 
  \lrfp{10.75}{g_{*d}}{\frac{1}{4}} 
  \lrf{10^{-7}\ {\rm GeV}^{-2}}{\la \sigma v \ra} 
  \lrf{m_\chi}{100\ {\rm GeV}} 
  \lrfp{1\ {\rm TeV}}{m_\s}{\frac{3}{2}}
  \lrf{\f}{10^{12}\ {\rm GeV}}
\end{eqnarray}
as long as the saxion decay provides a sufficient amount of the LSP. 

The initial LSP abundance produced from the saxion decay depends on
the saxion density at the decay as well as the branching ratio of the
gaugino production.  Since the saxion potential is generally flat, it
can develop a large expectation value during inflation like a modulus
field. When the Hubble parameter becomes comparable to the saxion
mass, it starts to oscillate around the potential minimum with an
amplitude $\s_i$. The cosmological abundance of the saxion thus
depends on the initial displacement from the potential minimum, which
is expected to fall in the range from $F_\A$ to $M_P$.
The saxion dominates the the energy density of the universe if the
following condition is satisfied:
\begin{eqnarray}
\frac{\s_i}{M_P} > \sqrt{6} \lrfp{T_d}{T_R}{\frac{1}{2}},
\end{eqnarray}
where $T_R$ is the reheating temperature, and we have assumed that the
reheating is not complete when the saxion starts to oscillate.  If the
saxion dominates the universe, the saxion-to-entropy ratio at the
decay is given by
\begin{eqnarray}
  \label{eq:ys-dom}
  Y_\s \;\simeq\; \frac{3}{4}\frac{T_d}{m_\s}
  \;=\; 8.5 \times 10^{-5}  \,\xi\,
  \lrfp{10.75}{g_{*d}}{\frac{1}{4}}
  \lrfp{m_\s}{1\,{\rm TeV}}{\frac{1}{2}}
  \lrf{10^{12}\,{\rm GeV}}{F_\A},
\end{eqnarray}
independent of the initial amplitude $\s_i$. 
The initial LSP abundance is related to $Y_\s$ as
 $Y_\chi(T_d) = N_\chi Y_\s$, where $N_\chi$ is the averaged 
 number of the LSP produced by the 
decay of one saxion field. Since $N_\chi$ is close to unity from 
Eq.~(\ref{eq:saxion_gauginoBr}), 
 $Y_\chi(T_d)$  easily exceeds the critical value $Y^{(c)}_\chi$,
and therefore the relic LSP abundance
saturates to the value given by (\ref{eq:relic_abundace}).

Even when the saxion field does not dominate the universe before its
decay, the LSP coming from the saxion decay tends to exceed the
critical abundance $Y^{(c)}_\chi$.  Since the reheating temperature is
then bounded above due to the gravitino
problem~\cite{Kawasaki:2004yh}~\footnote{
  Throughout this paper we assume that there is no source for the
  late-time entropy production~\cite{Lyth:1995ka} other than the
  saxion.
}, the saxion is likely to start to oscillate before the reheating
process completes. This is the case if $T_R < T_{\s}$, where $T_\s$ is
defined by
\begin{eqnarray}
  T_\s \;=\; 2.3 \times 10^{10}\,{\rm GeV} 
  \lrfp{200}{g_{*osc} }{1/4}
  \lrfp{m_\s}{1 \,{\rm TeV}}{1/2}.  
\end{eqnarray}
Here $g_{*osc}$ counts the effective number of massless degrees of freedom 
at $H=m_\s$. The saxion abundance is
\begin{eqnarray}
  \label{eq:ys-sub}
  Y_\s \simeq
  2.1 \times 10^{-11}\, 
  \lrf{1\,{\rm TeV}}{m_\s} 
  \lrf{T_R}{10^6\,{\rm GeV}} 
  \lrfp{\s_i}{F_\A}{2}
  \lrfp{F_\A}{10^{12}\,{\rm GeV}}{2}.
\end{eqnarray}
Thus even if the saxion does not dominate the universe,
(\ref{eq:ineqY})  is  satisfied for $\s_i \gtrsim \f$.
So, in the following, we adopt 
Eq.~(\ref{eq:omega_LSP}) to estimate the relic LSP abundance, assuming 
that (\ref{eq:ineqY}) is satisfied. Note that it is the large branching 
ratio into the gauginos that enables the saxion to non-thermally produce 
a right amount of the LSP DM, even if it does not dominate the universe.

\begin{figure}[t]
  \begin{center}
    \includegraphics[width=9cm]{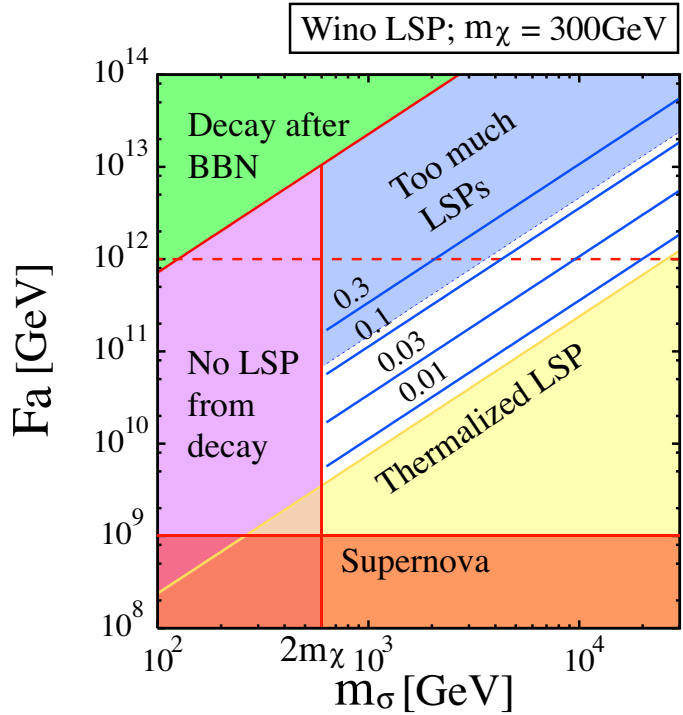}
    \caption{ Contours of the wino LSP abundance, 
      $\Omega_\chi h^2 = 0.3,\,0.1,\,0.03,\,0.01$, 
      for $m_\chi = 300$ GeV. 
      The shaded regions are excluded;
      the LSPs are only thermally produced and therefore too small
      relic density for $m_\s < 2 m_\chi$; 
      the upper left triangle region is excluded since the saxion 
      decays after the BBN starts, i.e., $T_d < 5$MeV, 
      while the lower right one corresponds to the decay 
      temperature higher than the decoupling temperature of the 
      wino, i.e., $T_d > m_\chi/20$; 
      $\Omega_\chi h^2 > 0.13$ is from the LSP overproduction in 
      the light of WMAP; 
      $\f<10^9$ GeV from the SN bound. 
      Also the region above the dashed line is disfavored by 
      the axion overclosure limit  if no entropy production occurs after
      the QCD phase transition. }
\label{fig:wino}
\end{center}
\end{figure}

In Fig.~\ref{fig:wino}, we show the contours of the LSP abundance 
in $(m_\s ,\f)$ plane. We have chosen the wino LSP with $m_\chi = 300$ GeV
and  $\xi=1$ as  representative values. 
Since the gaugino branching ratio is generically $O(1)$ as long as 
$m_\s > 2 m_\chi$, the relic LSP abundance is given by Eq.~(\ref{eq:omega_LSP}). 
From the observation of SN1987A~\cite{Raffelt:1996wa,Kolb:1990vq}, $F_\A$
is bounded from below, $\f > 10^9$ GeV, while an upper bound, $\f < 10^{12}$ 
GeV, is set by taking account of an overclosure limit of the relic axion 
abundance~\cite{Preskill:1982cy,Turner:1985si}. We find that the observed 
DM abundance, $\Omega_\chi h^2 \sim 0.1$, is realized for $m_\s$ around 
$O(10^{2-3})$ GeV and $\f = 10^{10-12}$ GeV (below the dashed line). Note 
that the axion overclosure bound can be relaxed by the entropy production 
associated with the saxion decay. In fact, in deriving the upper bound 
$\f < 10^{12}$ GeV, it is assumed that the initial displacement of the 
axion from the true minimum is $O(1)$, and that there is no entropy
production~\cite{Lyth:1995ka} after the axion starts to oscillate at
$T = \Lambda_{QCD} \simeq 200$ MeV. If the saxion dominates the universe 
and decays after the QCD phase transition (of course, but before the BBN 
begins~\cite{Kawasaki:1999na,Hannestad:2004px}), the upper bound can be 
relaxed to $\sim 10^{15}$ GeV~\cite{Kawasaki:1995vt}. Then although it 
depends on the initial VEVs of the axion and saxion fields, the region 
above the dashed line in Fig.~\ref{fig:wino} can be viable, and the right 
amount of the wino DM is realized even for that region~\footnote{
  The DM abundance includes the contribution from the relic 
  axion as well as the wino/higgsino LSP. When we take into account 
  the axion contribution, the smaller LSP abundance tends to be 
  favored, though the result depends on both the initial displacement of
  the axion and the thermal history of 
  the universe. 
}. 

In the above numerical analyses, we have fixed the wino mass as $m_\chi 
= 300$ GeV. In fact, the larger LSP mass favors smaller $F_\A$ for 
fixed $m_\sigma$. This can be understood 
as follows; higher saxion decay temperature is needed to realize 
$\Omega_\chi h^2 \sim 0.1$, since larger $m_\chi$ suppresses the annihilation 
cross section (see Eq.~(\ref{eq:ann_for_wino})). It is interesting to see 
that for $m_\chi \gsim 500$ GeV, the parameter region to reproduce $\Omega_\chi
h^2 \sim 0.1$ (calculated by using Eq.~(\ref{eq:omega_LSP})) overlaps with 
that where the LSP is thermalized, namely, $T_d \gsim m_\chi/20$. This means 
that the saxion decay takes place before the decoupling of the LSP. 
Then we need to numerically solve the Boltzmann equation around the LSP 
decoupling, taking account of the LSP production from the saxion decay 
in order to estimate the LSP abundance correctly. Since the relic 
abundance depends sensitively on the evolution around the decoupling 
temperature~\cite{Drees:2006vh}, we do not go into details in this paper. 
On the other hand, the lighter LSP may be excluded by the BBN. The $^6$Li 
can be overproduced due to the energetic hadrons from the DM annihilation. 
Thus a mass of $m_\chi \lsim 100 - 200$~GeV is disfavored for wino/higgsino
DM~\cite{Jedamzik:2004ip}. 

So far we have considered the wino LSP.
When the LSP is the higgsino, 
$\Omega_\chi h^2 \sim 0.1$ requires higher decay
temperature. In fact, since the annihilation cross section is lowered
by $\sim 1/10$, the decay temperature should be raised by about one
order of magnitude. Then the constant contours of the LSP abundance in
Fig.~\ref{fig:wino} shift to lower $F_\A$ by $\sim 1/10$ for fixed
$m_\sigma$. Therefore, we find that the observed relic abundance of
the higgsino DM can be naturally realized by the non-thermal
production from the saxion decay, for $m_\sigma$ at the weak scale and
$F_\A = 10^{9-12}$~GeV.

In the above discussions, we have considered the hadronic axion model.
Let us now comment on the DFSZ axion model. In this case the standard
model particles have nonzero PQ charges, and the saxion can dominantly
decay into the third generation quarks. Then the total decay rate of
the saxion becomes slightly higher and $N_\chi$ is suppressed by that
amount.  It should be noted, however, that the relic DM abundance is
given by Eq.~(\ref{eq:omega_LSP}) as long as (\ref{eq:ineqY}) is
satisfied.  Therefore the contours of the LSP abundance in
Fig.~\ref{fig:wino} slightly move upward for the DFSZ axion model, but
our main conclusion remains virtually intact.

Lastly let us comment on the gravitino production in the saxion decay.
If $\A$ has a mixing with the SUSY breaking field $z$, the saxion can
decay into a pair of the gravitinos, in addition to the single
gravitino production.  In contrast to the modulus,
however, the saxion decay is free from the gravitino overproduction
problem.  This is because the axion decay constant is much smaller than
the Planck scale.  Since the total decay rate is enhanced by lowering
the cutoff scale, the saxion mass needs not be larger than the gravitino mass
to decay before BBN, and 
the gravitino production may be kinematically forbidden. Even when the
channel is kinematically allowed, the branching ratio of the gravitino
production can be suppressed. Let us first consider the gravitino pair production
rate.  Since the coupling of $\A$ with the two gravitinos is proportional to
$F_\A$~\cite{Endo:2006tf}, the branching ratio of the gravitino
production is suppressed by $(F_\A/M_P)^4$, namely $\kappa_{3/2} =
O(F_\A/M_P)^4$. Since $F_\A$ is bounded above, $F_\A \lsim 10^{15}$
GeV, leading to $\kappa_{3/2} \sim 10^{-12}$ which is small enough
 to evade the bounds from the BBN and LSP overclosure
limits~\cite{Endo:2006zj}. However, the single-gravitino
production rate actually dominates over the pair production rate,
if the saxion mass is larger than $m_{3/2}$ due to the non-SUSY mass term.
The branching ratio of the single-gravitino production is given by
$O((m_\phi/m_{3/2})^2(F_\A/M_P)^2)$, and the cosmological bounds are
not so severe unless $m_\phi \gg m_{3/2}$.  Note that, since the saxion does not necessarily 
dominate the universe to produce a right amount of the wino/higgsino DM,
the gravitino abundance from the saxion decay may be suppressed.
Therefore the gravitino production can be safely neglected
in the saxion decay.

\section{Conclusions and Discussion}
\label{sec:5}

In this paper we have pointed out that an intermediate scale physics is
necessary for a successful non-thermal production of the wino/higgsino
DM, to avoid the moduli-induced gravitino/LSP problem recently pointed
out in Refs.~\cite{Endo:2006zj}. This is because 
gravitinos and/or LSPs would be overproduced otherwise.  The
conventional scenario using the modulus decay to produce the
wino/higgsino DM non-thermally is disfavored from this point of view.
Instead we have proposed an alternative candidate: the bosonic SUSY
partner of the axion, i.e., the saxion. It is particularly interesting
that the right amount of the DM can be realized for the saxion mass
around the weak scale when we consider the axion decay constant within
the cosmologically allowed range, $F_\A = 10^{9-12}$ GeV.  Although
our analyses  focused on the case of the saxion, we would like to stress
that they are rather generic and can be applied to any scalar fields
that decay via the non-renormalizable operators of $D=5$ with an
intermediate cut-off scale.

It is also worth noting that the saxion decay is
applicable for a source of the harmless entropy production. As was
stressed in this paper, the late-time decay of the scalar field generally
suffers from the gravitino/LSP overproduction. In contrast, the scalar
field at an intermediate scale can reheat the universe avoiding
overproduction of the gravitino/superparticles.  For instance, 
a scalar field at $M \ll M_P$ may decay before BBN begins,
while satisfying $m_\phi < 2 m_{3/2}, 2 m_{\rm LSP}$ to
kinematically block the gravitino/LSP production channels.

\section*{acknowledgement}

We are grateful to J.P.~Conlon and M.~Senami for comments. 
The authors would like to thank the Japan Society for Promotion of Science 
for financial support.

\end{document}